# A Quantitative Study of the Impact of Social Media Reviews on Brand Perception

A Thesis Presented to

the Faculty of the Weissman School of Arts and Sciences

Baruch College, The City University of New York

In partial Fulfillment of the Requirements of the Degree of

**MASTER OF ARTS**

**In**

**CORPORATE COMMUNICATION**

By

Neha Joshi

12/18/2015

# A Quantitative Study of the Impact of Social Media Reviews on Brand Perception

A Thesis Presented to
the Faculty of the Weissman School of Arts and Sciences
Baruch College, The City University of New York
In partial Fulfillment of the Requirements of the Degree of

**MASTER OF ARTS**

In

**CORPORATE COMMUNICATION**

By

Neha Joshi

12/18/2015

Under the guidance and approval of the committee,
and approved by all its members, this thesis has been
accepted in partial fulfillment of the requirements for the
Master of Arts in Corporate Communication.

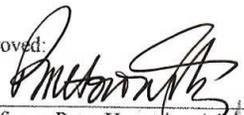
Approved:
Professor Peter Horowitz, Advisor    12/15/15 [Date]

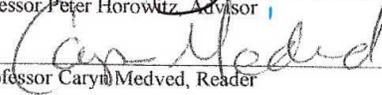
Professor Caryn Medved, Reader    12/16/15 [Date]

Professor Caryn Medved, Interim Graduate Director   [Date]

# A Quantitative Study of the Impact of Social Media Reviews on Brand Perception

## Dedication

Dad, your little angel did it! Mom, I did what it takes to complete my degree! Thank you to both of you. Your incredible support at crunch time and beyond cannot be expressed in few words.

A sincere thank you to my beloved husband, Dr. Aniruddha Marathe. I appreciate your support, encouragement, and sense of humor especially when I kept saying, "I don't know where I am heading with my thesis, and do I really have this under control?" Thank you for listening to this over and over, even at midnight!

I sincerely extend my gratitude to my maternal grandma. She was a woman of vision, dedication, and an immense love for life! Aji, I lived your dream. Grandpa, thank you for embedding in me a keen curiosity for knowledge.

A heartfelt thank you to my in-laws, uncles, aunts, cousins, and my closest friends. I dedicate this thesis to all of you.

# Acknowledgement

I would like to extend my most sincere appreciation for my advisor Professor Peter Horowitz. Thank you for believing in my thesis. I truly appreciate your valuable guidance, support, and words of encouragement. Your advice helped me focus on some of the most important contributions for this work.

I would also like to thank my reader Professor Caryn Medved. I sincerely treasure your most articulate and impeccable timely suggestions. Additionally, your interest in my work kept me motivated and keen to study the research problem more.

Thank you to my graduate advisor Cathy Levkulic. Your help was very crucial for a smooth Masters experience.

Thank you, Dr. Michael Goodman for all your support during the two years of my Master's degree. I would also like to take the opportunity to appreciate all my professors. Thank you also to all my friends at Baruch College, my Master's was much more doable because of you.

Finally, I would like to thank Jennifer Dessert of Endnotes Editing Services. Thank you for proofreading my work.

# Abstract

This thesis can be categorized under the *Influencer Marketing* industry with respect to social media initiatives. Influencer marketing is a modern tactic used by brands to enhance their visibility to their target audience by using the services of influential people. The objective of this thesis is to quantify the impact of social media reviews on brand perception. Specifically, this thesis focuses on two diverse media platforms commonly used for sharing opinions about products or services by publishing audio-visual or textual reviews: YouTube and Yelp. First, we quantitatively analyze the impact of YouTube reviews of Smartphones on the audience through their response to these video reviews. Second, using our findings from 942 YouTube Smartphone reviews, we introduce a statistical model to predict audience engagement on a given video. Finally, we apply our method of quantifying the impact of reviews on the content published on Yelp.com in the restaurant industry. The results from this validation show that our method can be generically applied to other social media platforms and consumer-focused industries. Our method can be employed by brand managers to turn social media reviews into real-time feedback mechanism in order to improve brand perception in the minds of their target audience.

## TABLE OF CONTENTS

# List of Figures

# List of Tables

# CHAPTER 1

# INTRODUCTION

Social media has grown beyond its original purpose of connecting people all over the globe. Today, the ease of interacting with millions of social media users in a cost effective and real-time manner has given rise to the creation of highly engaging online content by users from diverse professional and personal backgrounds. An all-encompassing definition of social media is "Social Media is a group of Internet-based applications that build on the ideological and technological foundations of Web 2.0, and that allow the creation and exchange of User Generated Content" (Kaplan & Haenlein, 2010 p.61). Successful social media platforms today are structured to incorporate a fast and cost-effective message delivery system that is globally accessible. Popular social media websites can be characterized by an ever-growing network of participants that interact through highly engaging content. Because of this ability to host persuasive content, social media platforms have become a powerful tool for brand managers to reach their target audience.

Recent advancements in technology, arts, and economics have greatly improved the usability and reach of social media platforms. For instance, a report by 2015 Pew research informs that there was a 7% rise in the usage of social media from 2005 to 2015. The report informs that 65% adults use social media (Perrin, 2015). As social media evolves into a more sophisticated tool for interaction and global reach, many individuals and companies are leveraging their influence using it to their benefit. As a consequence, the field of social media provides a viable career option for content creators who talk, sell, promote, and respond to audience effectively.

Keeping the competitive landscape in mind, the social media platforms are being leveraged not only by the content creators to maximize their reach to the target audience, but also by the companies selling the product being reviewed. To capture the interest of such a vast audience, a brand manager must measure the effectiveness of the channels that divulge brand information to that audience. In this thesis, we propose a formal technique to measure the effectiveness of these channels in conveying the brand message to their audience by using relevant user-relevant metrics.

Social media primarily attracts the attention of the millennial generation. According to the 2010 Pew Research report, the *millennial* is defined as having been born between 1977 and 1992 (Norén, L. 2011). The reviewers of the millennial generation have a high power of influence on the audience that thinks and acts like them. For the millennial generation, various social media channels have become inseparable part of every phase of their lives. Table 1 provides a broad classification of social media platforms (arranged in no particular order or feature preference):

TABLE 1

*CATEGORIES AND THEIR EXAMPLES OF SOCIAL MEDIA PLATFORMS*

| Type of Platforms | Examples |
|---|---|
| Text interaction and review | WhatsApp, Snapchat.com, Yelp.com |
| Personal space and entertainment | Facebook, Twitter, LinkedIn, Yammer.com |
| Audio and video | YouTube.com, 8tracks.com, Hubbub.com |
| Photography | Instagram, Flickr |
| Shopping | Amazon.com, BestBuy.com |
| Special topics | Runkeeper, interactive games, dating websites |

These platforms have enabled the millennial generation and others to become a part of their most personal interests and opinions. In this thesis, we specifically focus on two social media platforms used by the millennial generation: YouTube (video sharing) and Yelp (experience sharing).

The millennial generation is largely engaged in creating and viewing video content on popular video hosting websites such as YouTube. Due to the availability of inexpensive mobile and wireless internet to the millennial generation in the United States, the reach of YouTube content is enormous and the reaction of YouTube users to the videos is measureable. A comprehensive research by Statista.com shows that 53.6% of high-school graduates (graduated in June 2015) consider YouTube as the most favorite social networking website. In comparison, Twitter.com and Netflix.com attract 36.9% and 39.3% of the online population, respectively (Statista.com, 2015). According to Statista.com, the number of YouTube users in the United States has increased from 163.5 million in 2014 to 170.7 million in 2015, which is a significant increase of 7.2 million users per year. Statista predicts that the total users for YouTube will reach 187.8 million in 2019. These results complement with another insight that calculates the number of hours spent by people watching YouTube videos (Statista.com, 2015). According to YouTube.com statistics, "The number of people watching YouTube each day has increased by 40% y/y since March 2014" (YouTube Statistics).

Many reasons justify this growth in the total time spend on YouTube. One of the most crucial reasons is the advancement of YouTube from a channel that offers music to a full-fledged platform for online enthusiasts to upload opinion based video blogs or vlog on their own YouTube channels.

According to Oxford dictionary, a *vlog* is "A blog in which the postings are primarily in video form" (Oxford Dictionary). An individual using these vlogs are called vloggers. Vloggers belong mostly to the millennial generation and strive to attract maximum number of followers to their YouTube channels. Vloggers start their online presence as ordinary people. But, with the progression of time, some of them succeed in attracting a large number of views from the audience who can relate to the content in the videos over mutual interests. YouTube has helped ordinary people to turn into online celebrities in a short time.

Because of the way these vloggers have a powerful influence on altering the opinions of their followers, marketers can no longer ignore the possibility of employing these channels for brand promotion and reach. There is a professional aspect to vlogging that has evolved into a serious career choice for many of these vloggers that belong to the millennial generation. Because YouTube is so popular and provides a way to earn through creating and posting viral content, there is an opportunity to make a career as a video content creator. For example, The Guardian report states that, "YouTube star Felix 'PewDiePie' Kjellberg earned $7.4 million from his YouTube channel" (Dredge, 2015). Consequently, word-of-mouth promotion of popular brands has risen over years, especially after 2005 after the launch of YouTube.

While YouTube is a powerful tool, many reviews generate audience reaction in the form of comments that might be unrelated to the content of the video. To study the impact of reviews on the reaction of audience, this thesis focuses on an industry which might almost exclusively attracts audience engagement in the matter related to the content. Therefore, apart from a small number of unrelated comments, the industry selected for this study covers directly related reactions to the videos studied.

Within this framework, with 80 total number of channels, the technology industry is a popular market segment crowded by reviewers (YouTube). The fast-paced world of technology is seeing a rapid evolution in the products that substantially improve the quality of life of the users. For example, you can now experience augmented reality with Microsoft's HoloLens, which changes the way people perceive and interact with reality. The world of technology has always improved human-computer interaction to save time and draw the world closer.

For this thesis, we focus specifically on the smartphone reviews. The smartphone industry is a cluttered market with overwhelming information for potential buyers. Most buyers want their daily drivers[1] to have the best performance and style while being budget friendly. It is common place that smartphone companies launch new products or update their old products often, which is overly confusing for potential buyers when choosing a smartphone that fits their requirements. Smartphone buyers almost always rely on an advice from someone who has used their desired phone. Technology reviewers on YouTube offer their opinions about the features that the audience can relate to.

Because the users of technology products must be convinced about the effectiveness of features before making a buying decision, brand managers of technology companies must reach the most influential reviewers to convey the brand message of their product to their target audience. Empirically, research has been conducted on the impact of online content on the total number of views, subscription, and comments in isolation.

While these studies succeeded in measuring the impact of online content on websites like YouTube, they were limited in measuring the impact of content creators on YouTube on the audience in their full potential. For example, the total number of views taken in isolation may not be a sufficient indicator of the true potential of a reviewer's influencing capability. This is because views do not necessarily indicate audience retention and, therefore, do not categorize casual viewers from influenced audience.

---

[1] A daily driver is a popular term used by reviewers as a phone they use daily.

Research on the reaction of audience on YouTube must include a correlation between likes, dislikes, shares, and comments along with other factors for a truly inclusive measurement of brand performance.

This thesis uses a statistical technique to measure the correlation between attributes characterized by the techniques used by YouTube reviewers and the metrics indicating the response of the audience to these reviews. We use the following attributes (termed as 'variables' in the rest of the thesis) for our study:

TABLE 2

*ATTRIBUTES INDICATE TECHNIQUES USED BY YOUTUBE CONTENT CREATORS IN VIDEO PRODUCTION*

| Attributes |
|---|
| *Resolution of the video:* The size and number of pixels/sharpness of the video |
| *Tone of voice:* The quality of voice including intonations used by the reviewer |
| *Length of the video:* The total video time in minutes |
| *Published date of the video:* The date when the video was broadcast |
| *Channel:* Uploader's online space for collection of his/her videos |
| *Smartphone brand:* The technology companies of study |

TABLE 3

*REACTION OF THE AUDIENCE TO VIDEOS PRODUCED BY YOUTUBE CONTENT CREATORS*

| Metrics |
|---|
| *Views:* Number of distinct views for the video |
| *Likes:* Number of positive response to the video |
| *Dislikes:* Number of negative response to the video |
| Shares: Total number of shares of the video |
| *Comments:* Positive, negative or neutral response to the content of the video |
| *Subscription:* Number of followers of the channel |

Our research methodology can be explained in two parts: First, we classify video attributes crucial to the YouTube content creators and brand managers of the smartphone industry. We classify the video attributes as dependent variables (views, likes, dislikes, shares, comments, subscriptions) and independent variables (resolution of the video, tone of voice of the reviewer, length of the video, published date of the video, channel, and smartphone brand). Dependent variables enable us to understand the reaction of the audience to video attributes (independent variables) used by the reviewers to create the video. We use the well-known technique of simple linear regression to measure the correlation between the independent and the dependent variables.

This analysis demonstrates the brand perception and affinity before online content was produced and after it became viral. We present a linear regression-based model to predict the reaction of the audience (also called audience engagement in this work) on a video using only the attributes of the video in terms of *views-per-like* metric. We show that the predicted performance in views-per-like fairly follows the trend of actual performance on the videos of reviews on two smartphone brands: Nokia and BlackBerry.

Second, we apply our methodology by testing the correlation between brand performance in the restaurant industry and the reviews of diners on Yelp.com. We do not consider the reaction of the audience on the reviews for the restaurant since the intention was to manipulate the attributes to test the reaction of audience in the form of reviews to brand message delivered by companies especially because of its service nature. This is because, unlike the technology industry where reviewers would express their experience with the performance of the brand, in the restaurant industry the reviewer experiences the performance of the brand directly through the food and service (carried by the restaurant staff) in the restaurant outlets. The results from the second part of the method demonstrate that with minimal modifications, we were able to effectively apply the same methodology to the restaurant industry and quantify the impact of brand performance on the reaction of the audience.

The method used in our thesis is beneficial for both brand managers and reviewers particularly in the field of social media. Brand managers can use the method to understand the performance of the reviewers before investing enormous budget in on-line promotion. Brand managers may benefit from our study by filtering data that will help them make a more informed decision about using the service of the reviewers. On the other hand, reviewers can use our method for self-assessment to improve their performance, retain their followers, and work cohesively with the companies they review. Our method can work as a live mentor for the reviewers while they attract a meaningful audience engagement.

## Background of Study

The purpose of this study was to quantitatively examine the impact of social media reviews contributed primarily by the millennial generation on reaction of the audience to the commodities or services being reviewed. Specifically, this research quantified the way in which brand managers working for smartphone companies can leverage the services of highly influential YouTube content creators to reach a large set of online audience. Another key contribution of this work was a statistical model that predicts the impact of social media content, which can be beneficial to the communication team of a company. We described the usefulness of the statistical model in analyzing the interaction between a brand and its audience effectively for any industry other than smartphone. We demonstrated the usefulness of our impact quantification method by applying it to the social media reviews on Yelp.com focused on the Tex-Mex category of restaurants.

An increasing population of consumers is now relying on online reviews before buying a product or a service. This increasing trend compels brand managers to use social media platforms to promote their products and interact with their audience either directly or through popular influencers on social media. This research studied the content created by these social media reviewers, considered because of their popularity and ability to quickly make an online brand message viral.

Today, it is common to find companies using social media marketing as their primary strategy. One of the most impactful ways to communicate online is through audio-video messaging.

According to a report in The Guardian, "Nielsen claims 64% of marketers expect video to dominate their strategies in the near future" (Trimble, 2015). Companies focused on communication have social media experts (individual or group of professionals) who come together to plan, implement, and evaluate the social media initiatives in a time-sensitive manner. Marketers are willing to explore new techniques to improve their brand's social media presence to constantly keep up with the emerging markets. According to a recent report by Salesforce.com, "70% of marketers plan to increase their dollars for organic social or content marketing" (Heine, 2015). Consistent and integrated content marketing strategy is beneficial to the development of a strong relationship between a brand and the audience. Forrester Research defines content marketing as "a marketing strategy where brands create interest, relevance and relationships with customers by producing, curating and sharing content that addresses specific customer needs and delivers visible value" (Gerard, 2015).

Content creation on various social media platforms enables brands to promote visibility, audience interaction, promotion, reputation, and much more. Content creators are the catalysts of highly sharable brand advocacy messages as they guide their audience to generate opinions about these brands. Companies that use the services of influential reviewers get access to instant feedback and recommendations from their audience.

Various social media platforms facilitate easy and effective video content creation. YouTube provides content creators their online space in the form of channels that helps them publish free and high-definition videos on topics of their choice. Many research reports claimed that YouTube.com is the second largest search engine on the Web.

In a report, YouTube was shown as the most frequent video snippet domain compared to its competition (including hulu.com) (Edward, 2015). To help content creators stand out, YouTube provides several tutorials through their YouTube creator academy (YouTube Creator Academy) including lectures on high-definition visual designs, framing, lighting, and sound techniques.

Content creators use production attributes such as resolution, tone of voice, lighting, and sound to attract a large number of followers that respond to the video through metrics such as views, likes, dislikes, shares, comments, and subscriptions. Much research has been conducted on quantifying the views, subscription, and comments (in isolation) generated through a more generic set of videos. But no research has been conducted to quantify the correlation between video attributes used by content creators and the video metrics/reaction of the audience to these videos. Studying this correlation may help brand managers understand the technique of leveraging the services of content creators to improve the promotion of their brand.

In this research, we also examined the impact of brand performance on the reviews published on Yelp.com for the restaurant industry. On average, Yelp.com attracts 135 million monthly users (Wiideman). Yelp reviews benefit restaurant owners to interact with their customers, get feedback, watch their competitors and promote their popularity on Yelp.com. For this research, we focused on the Tex-Mex category of restaurants. We studied the reviews of Tex-Mex restaurants generated by the millennial generation located in a 10-miles radius of dense university campus areas in 8 major states.

## Framework

Content creators on YouTube start their career typically with a simple video using average quality video production equipment. As they gather subscriptions, content creators venture into advancing their production techniques to enhance the quality of their video. A well-designed plan to use the most advanced attributes for their YouTube content fetch them more followers. Companies looking to invest in online marketing primarily consider the popularity of these content creators as the deciding factor to approach them. In the smartphone industry, the strategy to promote their product on YouTube through these content creators is adopted at the earliest stage, typically even before the official launch of the product in their target market. Smartphone companies give out early editions of their upcoming phones to popular YouTube reviewers with a high subscription rate. While still in the pipeline, products reach those content creators that attract the maximum number of subscriptions and companies use additional analytics such as views and comments to decide which reviewer will benefit them the most. The performance evaluation techniques presented in our work can be readily adopted by brand managers to help leverage the attributes used by content creators to their benefit.

A good phone review are often produced with high-definition sound and visual quality since these attributes improve the effectiveness in detailing the visual features of a given smartphone through the video review. Use of correct lighting brightens up the room giving the phone a fresh look. The tone of voice used by the content creator helps them deliver a clear message. The length of the video syncs with the attention span of the audience, and the date the video was published justifies the reaction of the audience since technology is a time-sensitive industry.

In terms of these attributes, content creators have the power to influence the opinions of their audience about the brand and motivate healthy discussion about the phone features in the comments section of the video. The most popular metrics that show the reaction of the audience are views, likes, dislikes, shares, comments, and subscriptions. These metrics indicate the interaction of the audience with the brand and also the content creator.

A video *view* indicates a "conscious human choice to view a video". Views on YouTube are considered validated after the first 300 views on a video (Couzin, 2014). At this point, brand managers need to understand that considering the number of views in isolation does not cover the entire landscape of the reaction by audience. For example, one of the parameters used for this research is *views-per-like*, which shows number of views required for one like. Comments on YouTube could bring positive or negative response to video attributes that may influence fellow viewer/commenter. Likes and dislikes show audience loyalty and expectations respectively.

Content creators work as influencers by presenting their stories and opinions to their audience convincingly. Our research found that high production value results in higher subscriptions, which make content creators popular. These creators connect with like-minded followers who can understand the message in their content. Increased subscriptions motivate the content creators to constantly upgrade their methods and skills to convince popular brands to use their product review services before anyone else is given a higher priority.

Consequently, the audiences for these content creators feel a sense of privilege and belonging when connecting with these YouTube influencers and relate to their stories. Popular content creators on YouTube are considered celebrities and influencers by both the audience and the brand manager. Hence, studying the performance of these content creators is important. The next chapter explains the methodology used for this research.

# CHAPTER 2

# LITERATURE REVIEW

Understanding the impact of social media platforms in influencing opinions of the users has been an active area of research in the intersection of the fields of Computer Science and Marketing/Brand Management. This research was primarily focused on understanding how brand managers can leverage the knowledge of attributes of social media reviews to predict the popularity of their brand.

For YouTube, previous research was focused on modeling views, comments, and subscription to examine the popularity of video content in specific categories using sophisticated data mining tools.

## Diffusion of Innovation and Influence of Content

In their thesis *Examining the Diffusion of User-Generated Content in Online Social Networks*, Jeong et al. (Susarla & Tan, 2008) focused on diffusion of user generated content for social media influence for videos that are not genre specific. Diffusion of innovations is a theory that seeks to explain how, why, and at what rate new ideas and technology spread through cultures (Wikipedia). This research studied the impact of social media network among like-minded individuals on social learning and conformity. In contrast, our thesis examined the impact of content creators responsible for generating content conducive to expanding their social network. We studied specific attributes employed by content creators specifically on YouTube that enabled a successful social media network of like-minded individuals. Our research focused on the reasons for diffusion of innovation if considered in that framework.

## Study of YouTube Statistics

Research by Cheng et al. examined the correlation between videos published by content creators and algorithmically generated recommended videos by YouTube (Dale & Liu, 2008). While the research focused on the importance of content creator's choice, it emphasized the snowball effect by YouTube rather than the impact of these content creators on the choice by the audience. Additionally, this research also focused on measuring the total number of *views* generated by the videos as the most important metric of popularity. In contrast, as will be discussed later in this thesis, we prove that views alone cannot be a measurement of the popularity of the content creator because the total number of views do not cover influenced audience. Our research examined *views-per-like* and other parameters as crucial metrics to study the performance of the content creators.

## Comments and View Counts

Research was conducted on the impact of video content on view counts, comments, and video sharing for general categories by using linear regression and R model analysis to predict these parameters. A paper by Siersdorfer et al. focused on comments for published videos and meta-ratings for those comments. The research also analyzed sentiments in these comments using the sentiment-analysis tool, *SentiWordNet Thesaurus* (Siersdorfer et al, 2010). Further, this research predicted future community acceptance for comments that are not yet rated.

Richier et al. presented models to study the impact of the popularity of a video and its category based on evolution of the *view-counts* for the video (Richier et al, 2014). Their research focused on the views generated by YouTube users and how the prediction model can work to formulate the future view counts.

Haridakis and Hanson studied whether the motives (sentiments) of YouTube users can be used to predict viewing and sharing videos with others (Haridakis & Hanson, 2009). Their research was a quantitative study that was based on how audiences react to videos on YouTube. While these efforts involved analysis of the impact of videos on the response of the audience specifically for views, comments, and sharing; they did not study the correlation between these parameters. Furthermore, their work lacked a detailed examination of the reasons for the popularity of a video from the perspective of the content creator on YouTube.

## Prediction Model and the Importance of Content Creators

Very limited research had been conducted on the impact of the content creators on the audience engagement. In their work, *Broadcast yourself: understanding YouTube uploaders*, Ding et al. claimed that their research on the importance of content creators had never been done before (Ding et al, 2011). While they did focus on content creators, their work was not supported by substantially relevant quantitative analysis. Although their study offered a different approach to understanding YouTube reviews, they only presented a hypothesis.

While previous researchers focused on video length, video age (similar to date published that we studied for this thesis), and specific content (topic) discussed in YouTube videos with no specific focus on a genre or industry, current literature lacked a study of the attributes employed by content creators and how brand managers can leverage the knowledge of the techniques used by these content creators to capture audience engagement.

**Previous Work around Yelp.com**

In Chapter 3, we validate our methodology by analyzing restaurant reviews on Yelp.com. Previous research focused on sentimental analysis of YouTube reviews with a focus on positive and negative words (Hicks et al, 2012). However, previous work did not capture the performance of location specific outlets in close proximity to a specific target audience. Our thesis is unique in two aspects: First, we focused on specific locations within 10-mile radius of densely populated university campuses. Second, we specifically evaluated the performance of Tex-Mex restaurants across states in the United States with unique cultural diversities. Targeting outlet staff (carriers of restaurant experience in terms of food and services) and reviewers (restaurant diners) revealed unique insights that may help brand managers to efficiently invest their resources for a better return on investment.

# CHAPTER 3

# METHODOLOGY

This research used hypothesis testing to collect data that signified a relationship between dependent and independent variables. The type of investigation used was the correlation method. We used the interval scale within a linear regression model. An interval scale allows researchers to "perform certain arithmetic operations on the data collected from the respondent (Mir p.16)". We use this interval scale to measure the distance between two points on the scale. We collected 942 YouTube videos produced by tech content creators using the R language tool. These videos were published by 69 YouTube reviewers from United States and Canada for six smartphone brands. For the Yelp.com data, we collected 894 customer reviews for four Tex-Mex restaurants in eight states in the United States using Kimono labs software.

The rest of this chapter is organized as follows. First, we explain our method to gather review attributes on YouTube.com and Yelp.com. Second, we present our method to visually understand the correlation and interaction between various interesting attributes of reviews. Some interesting interactions between review attributes that arise in this step suggest potential for trivial as well as non-trivial correlations. Third, we present our attempt to formally quantify these correlations using the well-known method of Linear Regression. Finally, we present our method for applying the same method to quantify the performance of reviews on Yelp.com.

**Step I: Data Collection**

**Scraping YouTube Review Attributes**

To evaluate the reviews on YouTube, we divided the data in two parts: independent and dependent variables. We collected the independent and dependent variables, which are represented in Table 4 and Table 5.

TABLE 4

*VALUES ASSIGNED TO INDEPENDENT VARIABLES USED BY YOUTUBE CONTENT CREATORS*

| Attribute | Description |
|---|---|
| Resolution of the video | Value points for high quality (pixels) to low quality (pixels) videos |
| Tonality | Tone of voice with manually assigned values that indicate the intonation used by the speaker |
| Video duration | Calculated every 5 minutes of the total time of the video |
| Date published | Calculated as value point from assumed as Jan 01, 2016 |
| Channel | Space for collection of reviews by content creators |
| Smartphone brand | A balance of successful, underperforming, and a new brand |

TABLE 5

*VALUES ASSIGNED TO DEPENDENT VARIABLES THAT INDICATE THE REACTION OF THE AUDIENCE*

| Metrics | Description |
|---|---|
| Views | The total number of distinct views for the video |
| Likes | The total number of positive responses to the video |
| Dislikes | The total number of negative responses to the video |
| Shares | The total number of times the video was shared on other social media platforms |
| Comments | Positive, negative or neutral response to the content of the video that generates discussion |
| Subscription | The total number of followers to independent reviewers or reviewing company |

The smartphone brands include:

1. BlackBerry (Before Microsoft acquisition)
2. LG Electronics - Nexus
3. HTC Corporation
4. OnePlus (Before OnePlus two)
5. Nokia Corporation

**Scraping Yelp.com Review Attributes**

Kimono labs (Kimonolabs.com) software is a free web crawler that is used in this thesis to work with the Yelp.com reviews. A web crawler "is a program that visits Web sites and reads their pages and other information in order to create entries for a search engine index" (Rouse, 2005). We used Kimono labs to crawl Yelp.com reviews for the Tex-Mex restaurants in selective locations close to the university campus of selected cities and towns in the United States. Kimono labs use their technique to help users create their own application programming interfaces (APIs). These APIs are sets of requirements that govern how one application can talk to another (Proffitt, 2013). We could create our own API without writing a code, which was easy for us to understand data crawling and collection. The software is online, which makes it accessible through any device with the only requirement of an internet connection.

We registered for Kimono labs software with a login and password followed by installing a Kimono plug-in. On the main page, we copied and pasted the URL of the website that Kimono needed to crawl. After a few seconds, Kimono indicated that we needed to build our API page where we selected the required variables to be crawled. After Kimono labs scraped the data, we were instructed to save the API online, which we then transferred to an excel sheet for use.

Additionally, Kimono lab allows users to set a timer for continuous scraping of data on the selected websites. We used this data to input the values in R language and used linear regression to build a linear regression model. We observed that with minimum modifications, we could produce a linear regression model for Yelp.com reviews similar to YouTube data.

For the Yelp.com analysis we studied the correlation between the performances of four brands across universities in eight major states of the United States, as shown in Table 6.

TABLE 6

*LOCATIONS OF UNIVERSITIES FOR YELP.COM REVIEWS*

| Location | Universities |
|---|---|
| Alabama | Alabama A&M University, Alabama State University, and University of South Alabama |
| Arizona | University of Arizona |
| Connecticut | University of Connecticut and Trinity College |
| Hawaii | Hawaii Pacific University |
| North Carolina | North Carolina State University |
| Nebraska | Grace University, |
| Ohio | The Ohio State University |
| Texas | University of Houston, Texas Women's University, and Rice University |

We analyzed the performance of four brands in outlets within a 10-mile radius of these universities. These brands were Chipotle Mexican Grill, Moe's Southwest Grill, Qdoba Mexican Grill, and Taco Bell.

We collected over 10 years of data from 2005 to 2015 using the following dependent and independent variables for Yelp.com data:

TABLE 7

*DEPENDENT AND INDEPENDENT VARIABLES FOR YELP.COM REVIEWS*

| Dependent variable | Independent variable |
|---|---|
| Wordcount for Positive and Negative words | State Positive Words |
| Date published | Restaurant Date published |

**Step II: Understanding the Correlation and Interaction of Various data points for**

**Data Representation: Box Plots (also known as 'Error Plots')**

Box plotting is a commonly used way of representing the statistical characteristics of a set of discrete numerical values. These characteristics primarily include the mean, median, best-case and worst-case values of the data series being considered. Through box plots, one can derive correlation patterns in two or more data series. The graphical representation of data is called a box plot because it is a rectangle with lines dividing this box into groups of data called 'quartiles' (typically 25% of each quartile makes a box). The quartiles are *median, interquartile range, upper quartile, lower quartile*, and *whiskers*. Box plot shows the range of values on the Y-axis and categorization based on the other data series on the X-axis. Table 8 is an explanation of these components.

TABLE 8

*BOX PLOT DATA COMPONENTS*

| Component | Description |
|---|---|
| Median | The median (middle quartile) marks the midpoint of the data and is shown by the line that divides the box into two parts. Half the scores are greater than or equal to this value and half are less |
| Inter Quartile | The middle "box" represents the middle 50% of scores for the group. The range of scores from lower to upper quartile is referred to as the interquartile range. The middle 50% of scores fall within the interquartile range |
| Upper Quartile | 75% of the scores fall below the upper quartile |
| Lower Quartile | 25% of scores fall below the lower quartile |
| Whiskers | The upper and lower whiskers represent scores outside the middle 50%. Whiskers often (but not always) stretch over a wider range of scores than the middle quartile groups (Wellbeing@school). Box plot (or Error plot) is one way of seeing the spread of various data points to graphically represent the correlation between two data series. |

**Dependent and Independent Variables**

A variable is any given data series that is considered to be a component for statistical measurement. An independent variable is an attribute that describes the fundamental characteristic of the entity being described. On the other hand, a dependent variable is any variable that is influenced by the independent variable due to causality relationship. At this point, the change in the dependent variable is observed and recorded. For example, in this thesis the dependent variable of *views* on YouTube can potentially change (increase or decrease) with the change in the independent variable *duration of the video*, as we explore later. Independent variables are exclusively controlled by the reviewer. In formal scientific methods, independent variables are changed or controlled in the mathematical/scientific experiment to measure how they affect the dependent variables (Marie).

Any dependent variable is preferably represented on the Y axis and independent variables are likewise represented on the X axis to measure the impact of changing the latter on the former. In this thesis, we considered either one dependent variable at a time, such as views, or a combination of dependent variables such as views-per-like.

Table 9 lists the dependent and independent variables used in this study (smartphones and restaurants).

TABLE 9

*VARIABLES USED IN THE STUDY*

| Variable Type | Variables |
|---|---|
| Independent | Resolution of the video |
| | Tonality |
| | Video duration |
| | Date published - smartphones |
| | Date published – Restaurants |
| Dependent | Likes |
| | Views |
| | Views-per-like |
| | Views-per-share |
| | Shares |
| | Wordcount-per-positive |
| | Wordcount-per-negative |

**Positive and Negative Words in Yelp Reviews**

In case of Yelp.com reviews, we found two additional dependent variables for the completeness of the study: total positive words and total negative words in the review. The correlation between positive/negative words and other variables in Yelp reviews show direct correlation of the reaction of the audience to the brand performance and the brand message delivered by the restaurant staff through food and service in the outlets. Positive and negative words in the reviews are a direct feedback describing first-hand user's experience. Quantifying positive and negative words in the review itself is important because unlike YouTube, the interaction on Yelp.com is directly between the brand and the user of the service.

**Step III: Modeling the Performance of Video Reviews from Independent Variables**

**Using Linear Regression**

In their book *Introduction to Linear Regression Analysis*, authors Montgomery, Peck, and Vining define linear regression as "a statistical technique for investing and modeling the relationship between variables" (Montgomery et al, 2013). Regression-based models are used extensively by statisticians to solve real-life problems in many fields including management and other social sciences. Linear regression is a mathematical technique used to analyze how change in one variable can affect the other variable.

In this thesis, we use linear regression to analyze the following:

1. Correlation between the attributes used by content creators on YouTube.com for smartphone companies and the reaction of the audience to those videos measured through YouTube audience metrics. For example, we try to find the correlation between resolutions used by content creators and likes generated by the audience on that video.

2. Correlation between the performance of the brand and the reaction by the audience in form of reviews on Yelp.com for Tex-Mex restaurants. For example, we try to find a correlation between the positive words and the date published of the review. Date published is an important attribute to understand the age of the video and the trend of influence over time.

Linear regression is a commonly used statistical method used to measure the correlation between two (related or unrelated) data series. Although the correlation may not necessarily indicate causality[2], quantifying the correlation is crucial to understand the impact of important characteristics of the data sets being compared. We present techniques to apply linear regression to the social media content attributes and properties of the reaction of the audience/metrics. This study may help brand managers understand their customers and the performance of their brand most effectively.

Results delivered with the use of the linear regression technique illustrate a possible relationship between two variables typically illustrated with a scatter plot. "Scatter plots are used to plot data points on a horizontal and a vertical axis in the attempt to show how much one variable is affected by another" (Spotfire.com). For this thesis, we use a box plot instead of scatter plot to simplify the visual aspects of the results.

**R Language to Apply Linear Regression Model**

R is a computer programming language in a system for 'statistical computation and graphics' (Becker, Chambers & Wilks, 1988 p. 1). R language helps statisticians and data scientist to find astonishing insights into the data by simply plugging the acquired numbers in the R system. R is an open-sourced language available to download for free and many online tutorials are available on how to use R most effectively to transfer raw data into graphical representations. In his article "Big Data: What is R? R Explained in less than Two Minutes, to Absolutely Anyone," Bernard Marr of Advanced Performance Institute explained that

---

[2] Casual analysis in linear regression determines whether change in independent variable really affects dependent variable.

R's strengths as a statistical programming language draw from the fact it is designed from the ground up to facilitate matrix arithmetic - carrying out complex, often automated calculations on data which is held in a grid of rows and columns. R is very good for creating programs which can carry out calculations on these datasets, even when the datasets are constantly growing in size at an ever-increasing rate, and producing real-time visualizations based on this data (Marr).

Marr further stated that R helps statisticians represent their data graphically through "charts, graphs and complex multi-dimensioned matrices." Through this thesis we found that statistics is incomplete without programming. R programming language is important for businesses because of its open-sourced nature. Additionally, transferring big quantities of data from an excel sheet (collecting and adding data to the excel sheet as the first step) to graphs can be tedious and expensive for companies trying to understand the impact of their business model and the framework of the reaction of their audience (Amirtha, 2014). In this thesis, we use the R language to (a) transfer dependent and independent variables into combinations of several graphical results, and (b) to predict a linear regression model that shows changes to the dependent variable with modifications in the independent variable for the future YouTube videos and Yelp reviews.

**Linear Regression-Based Model to Predict Views-per-Like**

Linear regression model is provided as a readily available tool in the R language. The components of linear regression model are as follows.

**Best fit line.** The best-fit line is the 'mean' of the data. Mean is calculated by adding all the values of the data points and dividing the total by the number of data points.

**Residual.** A residual is the distance from the mean to the observed value on a graph. Simply put, residuals also called 'errors' are values that are scattered on a graph plot at a position that is near or far away from the mean of the data or best-fit line. When the residuals are added they always result in zero. Residuals are squared to make the distance positives and to emphasis a larger deviation (similar to standard deviation). The squaring exaggerates the points that are further away. The goal of simple linear regression is to create a linear model that minimizes sum of squares of the residuals (Foltz, 2013).

**Coefficients.** These are the constants (value that do not change) that give you the best case estimation of predicted values using the model. The use of this model and the actual model will be explained in the next chapter.

# CHAPTER 4

## DISCUSSIONS AND FINDINGS

The data is divided into two parts. (a) Attributes used by content creators to publish a video and (b) the metrics used for measuring the reaction of the audience to the videos (see Table 4 and Table 5). As discussed earlier, the primary purpose of this thesis is to find a correlation between the dependent variables and the independent variables for 942 YouTube videos.

While views and likes are important metrics to measure the reaction of the audience to videos, our work found that they are not sufficient to understand the true potential of what is preferred by the audience. Views, for example, do not capture audience reactions, i.e. views are merely the total amount of validated time spent watching a video, but does not include the total number of influenced viewers. While likes is a better parameter to show the positive response to the videos, we found that likes in isolation does not cover viewership. For example, consider a video received 1000 views and only 100 likes. The content of this video clearly failed to attract maximum positive reaction from the audience. Through this result, we found that views-per-like metric captures the influence of content on the audience more effectively than other candidate composite metrics. Consider the following analogy: views-per-like are similar to miles-per-gallon calculated to assess the performance of a car in terms of mileage. Through views-per-like, we essentially capture the mileage of the videos.

We studied the correlation of views-per-like and other variables for six smartphone brands. We picked three contrasting brand categories: old and successful, old and unsuccessful, and new brands in the smartphone industry. The reason for this selection was to identify the impact of YouTube reviewers on audience with content of varied brands with varied backgrounds and experience in the industry. Table 10 outlines the video attributes of the following reviewers/channels we examined.

TABLE 10

*YOUTUBE VIDEO REVIEWERS AND CHANNELS STUDIED*

| Reviewer | Channel |
|---|---|
| Aaron Baker | PhoneDog |
| Austin Evans | Austin Evans |
| Brandon Miniman | Pocketnow |
| Chris Pirillo | Chris Pirillo |
| Danny Winget | Danny Winget |
| Erica Griffin | Erica Griffin |
| Flossy Carter | Flossy Carter |
| John Velasco | PhoneArena |
| Jon Rettinger | TechnoBuffalo |
| Joshua Vergara | Android Authority |
| Linus Sebastian | Linus Sebastian |
| Lisa Gade | Mobile TechReview |
| Marco Hanna | PhoneDog |
| Marques Brownlee | MKBHD |
| Michael Fisher | Pocketnow |
| Michael Kukielka | DetroitBORG |
| Sassy Tech Girl | Sassy Tech Girl |
| Todd Haselton | TechnoBuffalo |

We carefully monitored the performance of these content creators for the brands discussed in the Data Collection section.

As mentioned earlier, our research measured the influence of smartphone reviewers on YouTube based on selective attributes and metrics. Since the R language can operate primarily on data series consisting of numbers, we assigned values for our attributes. We measured the impact of views-per-like on resolution of the video and tone of voice[3] (also called tonality in this thesis) used by the content creator. Table 11 presents the values that were provided to the R program for resolution:

TABLE 11

*VALUES ASSIGNED TO THE R PROGRAM FOR RESOLUTION*

| Maximum Resolution | Values Assigned |
| --- | --- |
| 2160/4K | 8 |
| 1440 | 7 |
| 1080 | 6 |
| 720 | 5 |
| 480 | 4 |
| 360 | 3 |
| 240 | 2 |
| 144 | 1 |

---

[3] In this thesis, we use tone of voice and tonality interchangeably

Our study found that a high-resolution video is pleasing to the eye and can immediately bring a sense of reliability. For example, Figure 1 shows a snapshot of a video with a low resolution of 240 pixels and Figure 2 shows a snapshot of the same video with the maximum resolution of 2160 pixels (also known as 4K). In these figures, the difference in the pixels and hence the difference in quality of the picture is evident. Our study found that the resolution is crucial in audience engagement and retention.

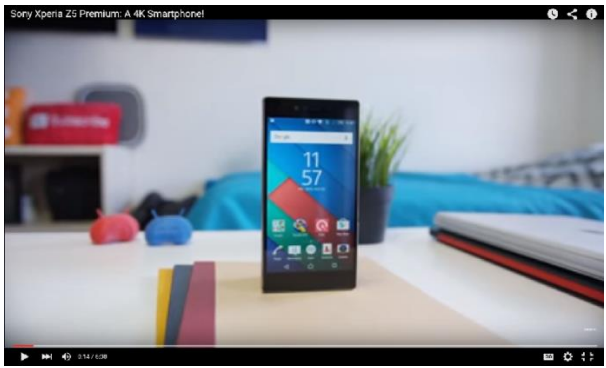

*FIGURE 1.* 240P LOW RESOLUTION

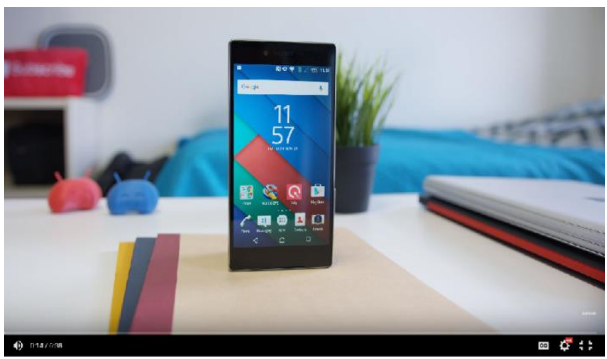

*FIGURE 2.* 2160P/4K HIGH RESOLUTION

The next important variable is the tone of voice used by the content creators. Tone of voice or 'tonality' is characterized by the intonations and pitch used in a video produced by the content creators. As mentioned earlier, since R program understands numeric values, we assigned assumption values to categorize tonality. We manually watched every video for the first 60 seconds and noted down the most relevant category of tonality. This was based on the sharpness in the voice and the intonation used by the speaker. We found that the sharpness in the voice is a very important element in not only attracting the audience but also sustaining them as followers. On the other hand, the quality of intonation makes the content sound either interesting or dull.

We assumed the following categories and values to those categories, which are reflected in Table 12.

TABLE 12

*VALUES ASSIGNED TO TONALITY OF SPEECH*

| Manually Assigned Categories | Values | Effective Quality |
|---|---|---|
| High Crispness/clarity - High intonation | 10 | Good |
| High Crispness/clarity - Normal intonation | 9 | Better |
| High Crispness/clarity - Low intonation | 8 | Average |
| Normal Crispness/clarity - High intonation | 7 | Average |
| Normal Crispness/clarity - Normal intonation | 6 | Good |
| Normal Crispness/clarity - Low intonation | 5 | Bad |
| Low Crispness/clarity - High intonation | 4 | Bad |
| Low Crispness/clarity - Normal intonation | 3 | Average |
| Low Crispness/clarity - Low intonation | 2 | Bad |
| Not audible | 1 | Worse |

We also studied the correlation between the *date published* attribute and views-per-like. For simplicity, we calculated *date published* backwards by quarter, i.e. the number of days between the date published recorded on a video and January 1, 2016 (alternatively, each quarter from January 1, 2011) and then grouped the videos according to the quarters in which the new date published fitted. This helped us understand the trend of the video over time since it was published first.

As outlined in the previous chapter, our methodology also included a linear regression based model to predict the performance of the video in terms of views-per-like. Table 13 provides components of the prediction model.

TABLE 13

*COMPONENTS OF THE PREDICTION MODEL*

| Parameter | Description |
|---|---|
| Formula | To establish a correlation between views-per-like (dependent variables) and the independent variables |
| Residuals | Best-case error between the actual and the fitted values of views-per-like |
| Coefficients | These are the constants (value that does not change) that give you the best case estimation using the model. |
| Views-per-like | $(-1.051e+00)$ x tonality + $(-4.190e+00)$ x resolution + $1.565e+01$ x video duration + $5.456e-03$ x date published + $(-6.326e+01)$. |

The findings of our YouTube study are discussed in the next section of the chapter. We validated our method by applying it to the reviews for Tex-Mex restaurants on Yelp.com. Just like views-per-like for YouTube reviews, we studied wordcount-per-one-positive-word for Yelp.com. Similar to views-per-like, this metric focused on the number of positive words numerically divided by the total word count/number of words in each review. For example, 15 positive words in a review with 20 wordcount makes it a powerful review compared to a review consisting 5 positive words per 20 total number of words. We present our findings in the next section.

**Findings**

Our research findings give brand managers a guideline to use the potential of content creators to the fullest. Through the graphs presented in this section, we demonstrate that brand managers can delegate work of gathering data and mapping the results to be used for their study before any change in branding strategy. As a brand manager, carefully reading the outcome of the data gathered enables an unbiased and filtered selection of the best scenario to reach their audience most effectively. This section examines a few graphs generated by our method and guide brand managers with a way to understand and read these data carefully.

**Findings on YouTube.com**

    **Views-per-like by resolution.**

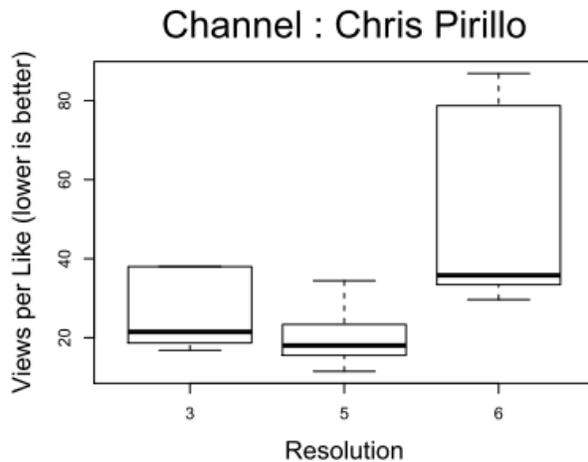

*FIGURE 3*. VIEWS-PER-LIKE BY RESOLUTION

*Contradictory results.* As discussed earlier, a high-resolution video results in positive audience engagement. However, we found a few baffling graphs that challenged this claim.

For instance, in Figure 3, though Chris Pirillo (a reviewer) used the highest available video resolution of 1080p, the response he received from the audience diminished in views-per-like. Views-per-like is the total number of views required to receive one like, and show the total number of views converted into influenced audience (based on likes).

We observe that the highest resolution of 6 (1080p) generates high views-per-like. We investigated this graph further by comparing it with a correlation between views-per-like and duration of the Chris' videos as seen in Figure 4. We estimated that the reason for high views-per-like at high resolution of video (bad result) might be a result of longer videos produced at the 1080p resolution. Our estimation was close to the results generated after analyzing further.

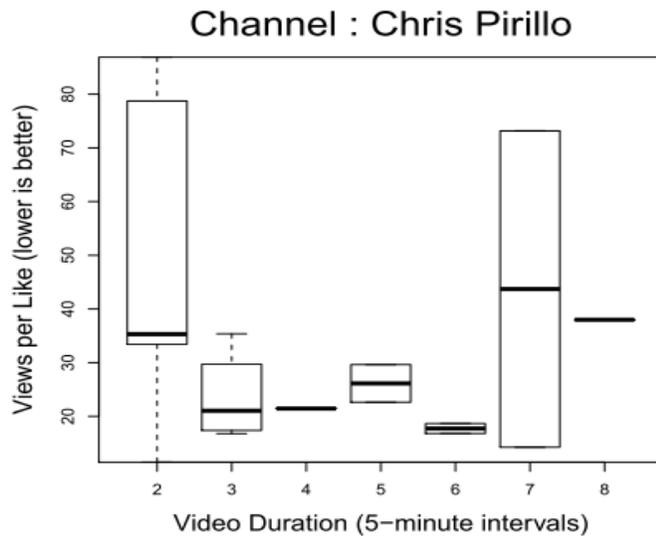

*FIGURE 4.* VIEWS PER LIKE BY VIDEO DURATION

On investigating further, we found that the worst performing video for Chris Pirillo is between 35 and 40 minutes (see points 7 and 8 on the bar graph, Figure 4). Though we were getting closer to our finding, the results were still redundant. At this point, we manually watched these videos and noticed (a) most of these videos used a resolution of 6 (1080p) and (b) Chris Pirillo called these videos "live" videos. We found that this was the cause of Chris Pirillo's audience losing their attention and interest in his videos in spite of a high resolution.

Live YouTube videos are unedited and longer videos produced by content creators. We found that producing a live video could be a good step towards high interaction with the audience only if the video does not stay longer in the content creator's channel. This could be because the viewers of a live video might not watch the video live but much later. But, because the video was aired live, it will have unedited content susceptible to flaws.

To prove that this observation is true, we looked at two supporting graphs that included (a) likes by date published (See Figure 5) and (b) views by date published (See Figure 6).

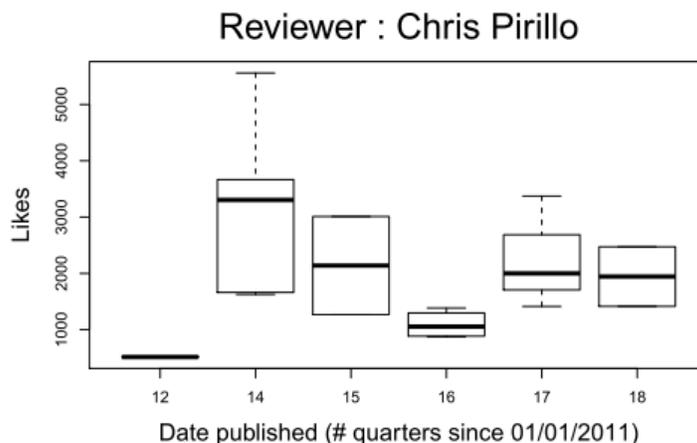

*FIGURE 5.* LIKES BY DATE PUBLISHED

Notice in Figure 5 the first quarter Chris Pirillo's videos generate less than 1000 views. In the second quarter, the total number of views increased dramatically followed by a bumpy curve. However, there was a significant increase in the views generated in Quarter 17 (notice a very high rise).

On further investigation we found that as seen in Figure 6, increase in the total number of views did not affect the total number of likes (See Figure 5), i.e. there was no significant improvement in the likes especially for the most recent videos.

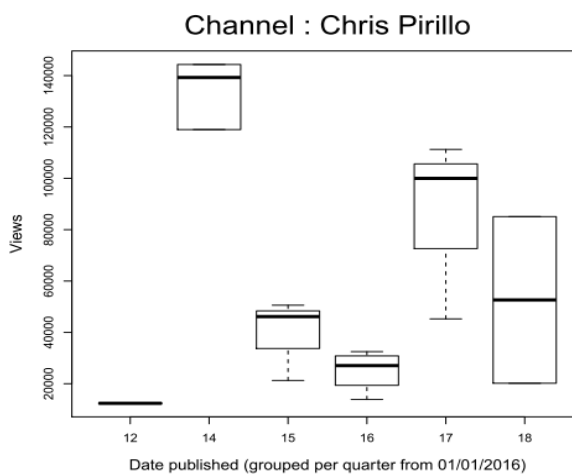

*FIGURE 6.* VIEWS BY DATE PUBLISHED

Based on the data collected, we noticed that the most recent videos were mainly unboxing videos, longer and live videos. The results indicate that: (a) To be successful, a good unboxing video must be short and focused on unboxing and should cover relevant details about the features of the smartphone reviewed (b) A reviewer should not keep a live video on his channel for a long time because it might hurt the reaction to their performance for two reasons (i) viewership to live videos can increase much after the video goes live (ii) because it is live, the video is an unedited version i.e. could lack clarity of thoughts, expression, and resolution.

Hence, brand managers must look at views-per-like and dwell deeper if they see an unusual result to make a more informed decision about the choice of a reviewer. They must also remember that every type of video generated by a reviewer must follow certain rules to attract an unbiased and unfiltered attention of their audience. To further investigate the impact of a poor choice by a reviewer on the reaction of the audience, we examined videos by Sassy TechGirl.

**The Impact of Resolution on the Popularity of the Reviewer**

To establish our findings about the correlation between views-per-like and resolution and also to examine the impact of the decision taken by reviewers on audience engagement[4] we studied another case. Observe Figure 7, the views-per-like generated by videos with a low resolution are around 150 whereas, those generated by a higher resolution are 800. This shows that, although Sassy TechGirl improved the resolution of her videos, her audience did not accept the change. This proves that, Sassy Tech Girl's popularity diminished over time. Her service would only hurt the brands she reviews. We studied supporting drafts such as tonality of speech, video duration, and date published.

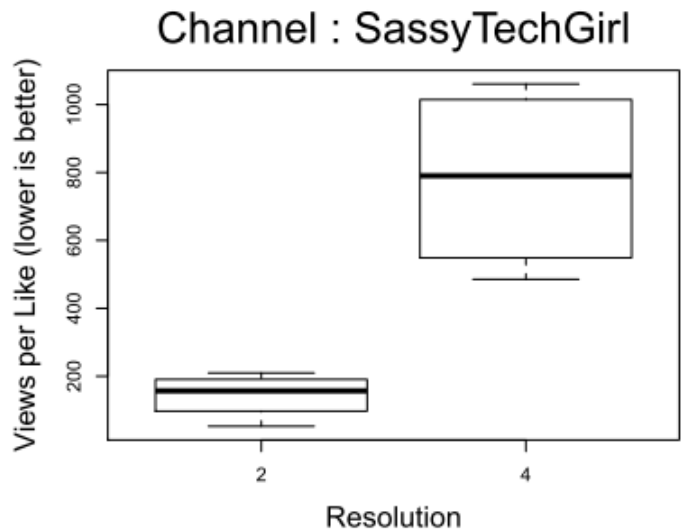

*FIGURE 7.* VIEWS-PER-LIKE BY RESOLUTION

---

[4] We use reaction of the audience and audience engagement interchangeably.

As seen in the figure 8, observe that Sassy Tech Girl makes use of a consistent tone of voice i.e. a low tonality of 2 (assigned the value that indicates bad tonality) i.e. low crispness/clarity and low intonation of speech.

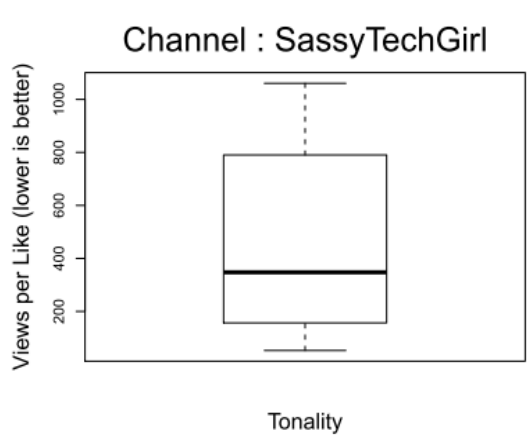

FIGURE 9. VIEWS-PER-LIKE BY TONALITY

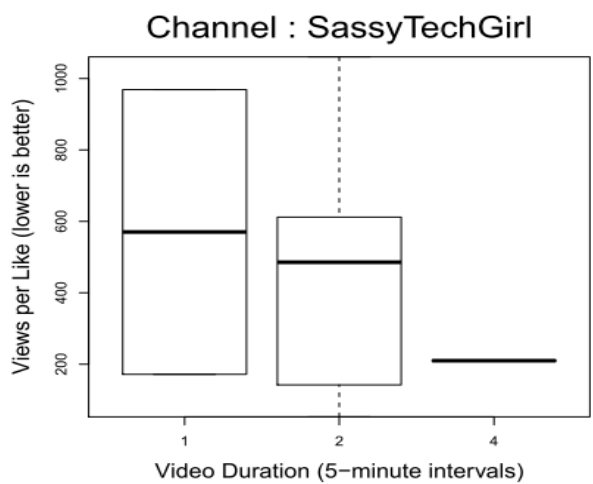

FIGURE 8. VIEWS-PER-LIKE BY VIDEO DURATION

However, as we investigated in-depth, we found that with changes in the attributes used, YouTube reviewers can improve their performance, hence reaction to their videos.

We found that the potential solution for Sassy TechGirl to improve the response of her audience to her content could be (a) improving her tonality and increasing the duration of her videos. We based this by observing views-per-like by video duration (See Figure 9) and date published by video duration (See Figure 10).

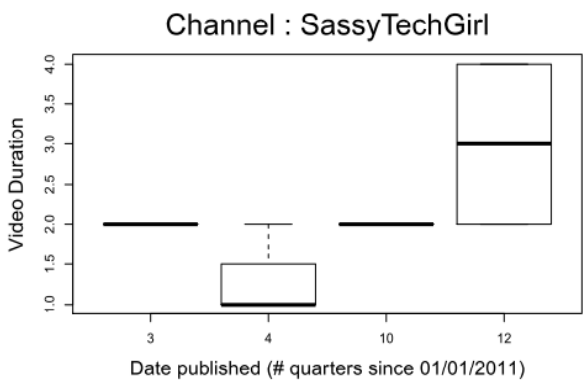

FIGURE 10. VIDEO DURATION BY DATE PUBLISHED

These figures illustrate a positive audience reaction to a set of 20-minute videos produced in the 12th quarter. This could indicate that a longer video duration that covers more features with a much better tonality could be the solution for Sassy TechGirl to achieve a good set of followers and positive reactions in the future. Sassy TechGirl is a very important example of the impact of YouTube production attributes on the performance of the reviewer and the impact of that performance on the reaction to the videos.

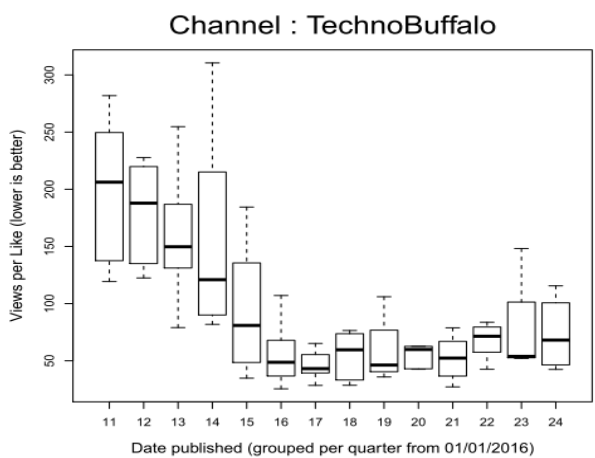

FIGURE 11: VIEWS-PER-LIKE BY DATE PUBLISHED

**Views-per-Like by Date Published**

For an industry as time-sensitive as that of smartphones, it was crucial for us to examine the correlation between views-per-like and date published. Our case study was a company that consisted of four reviewers, two of which made some impact on audience engagement.

We observed the performance of TechnoBuffalo in terms of views-per-like by date published. This was an interesting case since that provided us with two findings:

First, the performance of TechnoBuffalo videos was not good in the first quarter and it continued its downturn. In short, the views were high in the quarters between 11 to 13 and then dropped drastically. We considered that this could be because of the performance of the reviewers. To understand this further, we studied individual reviewer performance of Jon Rettinger and Todd Haselton of TechnoBuffalo.

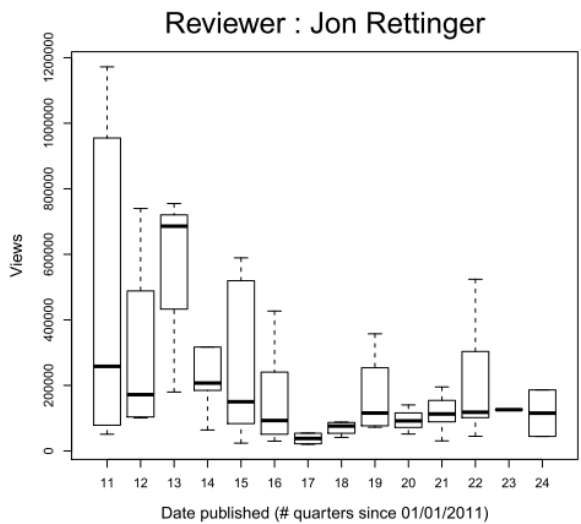

*FIGURE 12*. VIEWS BY DATE PUBLISHED FOR JON RETTINGER

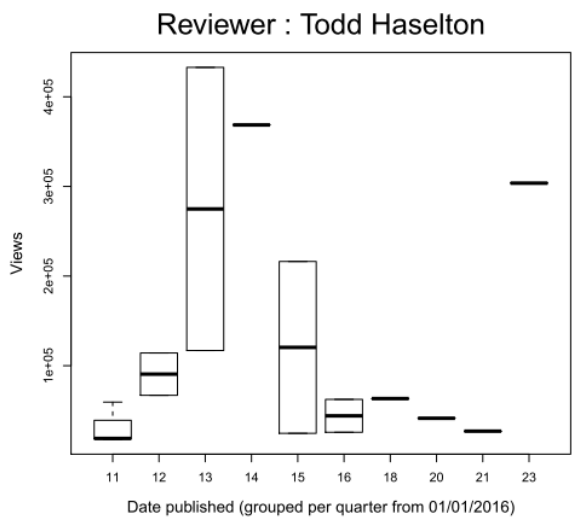

*FIGURE 13*. VIEWS BY DATE PUBLISHED FOR TODD HASELTON

We further investigated these reviewers to understand the impact of their performance on the response of the audience. Note Figures 13 and 14, we observed that Jon had higher views compared to Todd. However, he also had fluctuating likes and higher dislikes for those views.

We also observed that Jon received more comments (See Figure 14) compared to Todd (See Figure 15). But, because Jon received subsequently higher dislikes (See Figure 16) inconsistent likes (See Figure 17), there was a possibility that these comments could be inclined toward criticism of his content. It could be a possibility that the highest views-per-like in the 11$^{th}$ quarter could be because of an unsatisfactory performance by Jon. This could also mean that, in spite of a company publishing videos by their trained and hired staff, the audience might react to the reviewer independently and that the interaction between the reviewer and audience, both through his content and his response through videos that followed, could affect the reputation of the company that reviews brands on YouTube. While there is a limitation in proving this claim without analyzing the content and sentiments in the comments through more sophisticated analytical tools like Natural language processing, our study provides a direction for brand managers to conduct further research based on some guidelines that we provide.

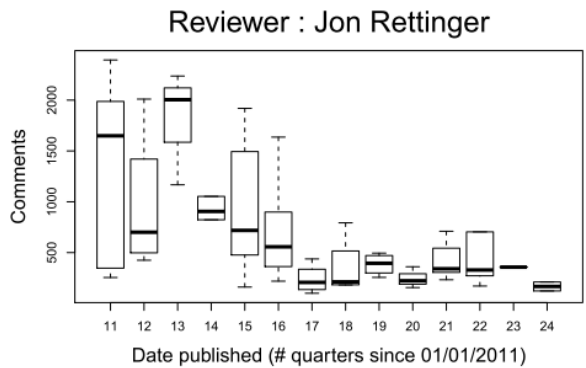

*FIGURE 14.* COMMENTS BY DATE PUBLISHED FOR JON RETTINGER

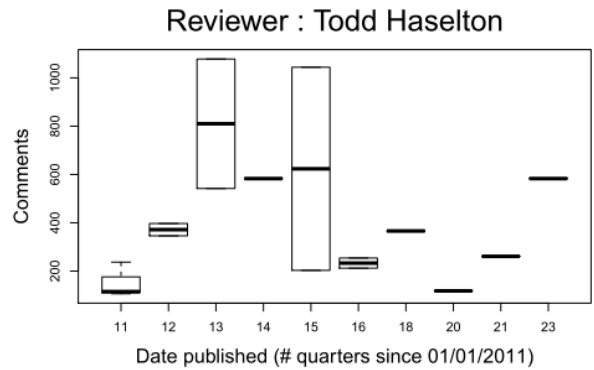

*FIGURE 15.* COMMENTS BY DATE PUBLISHED FOR TODD HASELTON

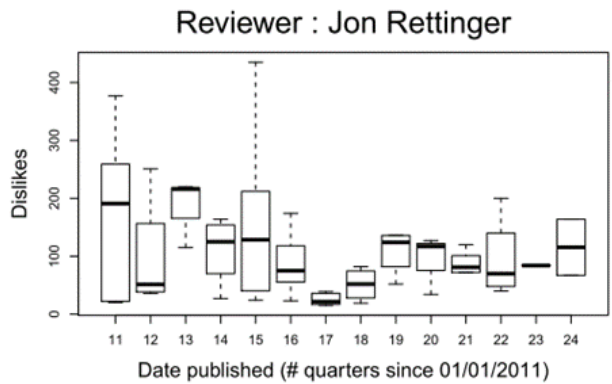

*FIGURE 16.* DISLIKES BY DATE PUBLISHED FOR JON RETTINGER

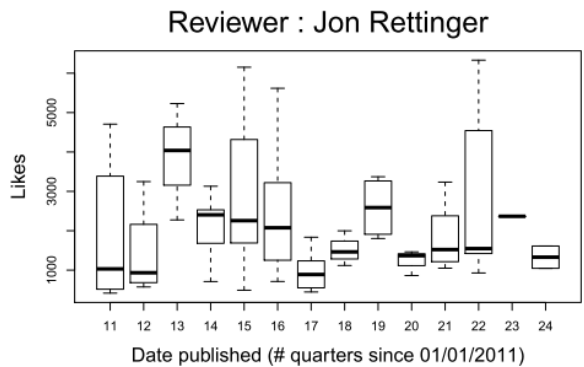

*FIGURE 17.* LIKES BY DATE PUBLISHED FOR JON RETTINGER

**The Impact of Popularity of a Brand on Positive Audience Engagement**

TechnoBuffalo's case opened a broader question for us. Can a brand play an equally important role in getting the attention of people towards a reviewer/video? Consider Figure 17: observe a sudden rise in the 13th quarter for likes in Jon Rettinger's video. We noticed this exact rise at the 13th quarter for almost all the reviewers in terms of attributes taken in isolation (Views, likes etc…). On looking closer, we found that this was the time when almost all the reviewers were reviewing Samsung Galaxy Note II. As we examine the graphs and watch some videos, we notice that the reason for a sudden spike was the brand itself.

However, our study indicates that some reviewers did not attract positive feedback during that span through Samsung Galaxy Note II reviews. This shows that brand managers need to be aware of their choice.

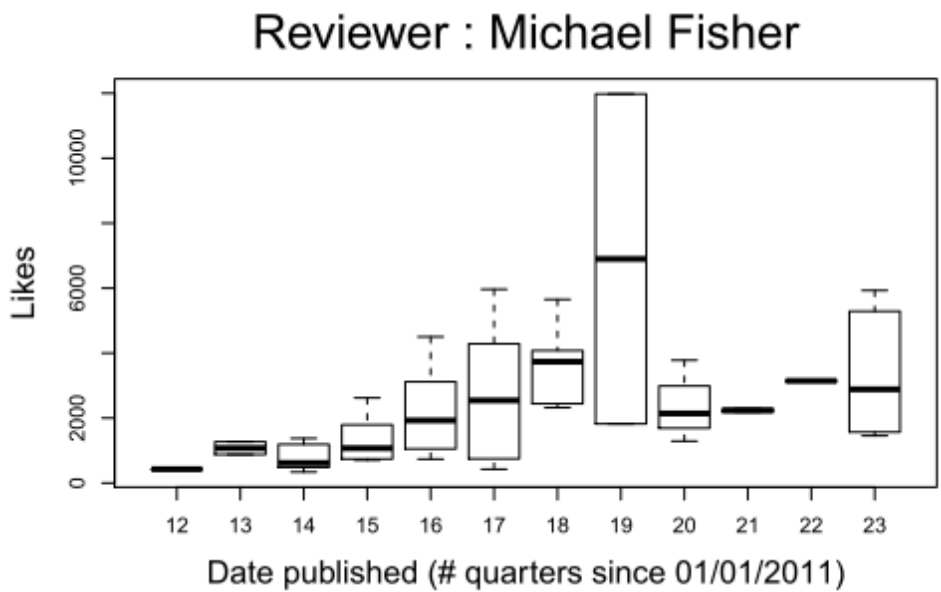

FIGURE 18. LIKES BY DATE PUBLISHED FOR MICHAEL FISHER

As stated above, observe Figure 18 where Michael Fisher is a contrasting example. We see a spike in the 19th quarter for videos produced by Michael Fisher. This may or may not be because of a case similar to Samsung Galaxy Note II reviews.

The crucial point to draw here is that there was a spike in the 19th quarter only for Michael Fisher's videos. This could also be because probably he caught high attention of the audience through his content or his interaction with the audience during that time. Hence, brand managers need to be aware that in a case like this, comparison between reviewers will help them make a much informed decision about the choice of their reviewers. By using a prediction model that we propose in this thesis, brand managers can now compare the actual values with the predicted values to understand the performance of the reviewers and use them to their best practice.

**Successful case study: OnePlus One**

      We added *OnePlus* much later to our study. The reason to include *OnePlus* was to analyze the impact of independent variables in videos for a new brand on the reaction received. For example, the advancement in video capture technology (i.e., video resolution) and its use in product reviews may lead to different trends in the performance.

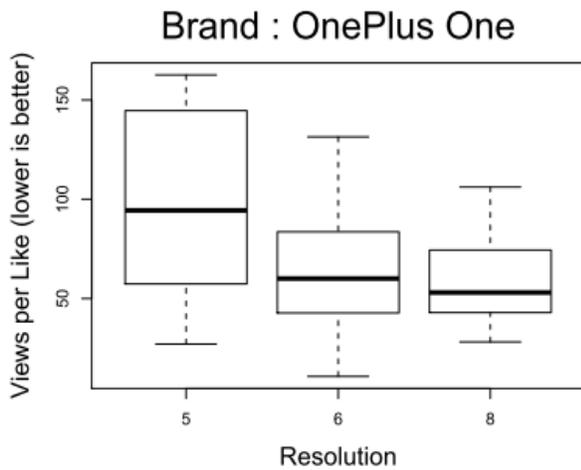

*FIGURE 19.* VIEWS-PER-LIKE BY RESOLUTION

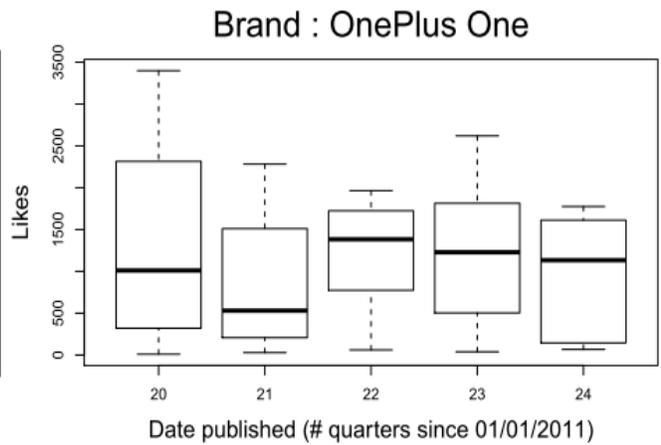

*FIGURE 20.* LIKES BY DATE PUBLISHED

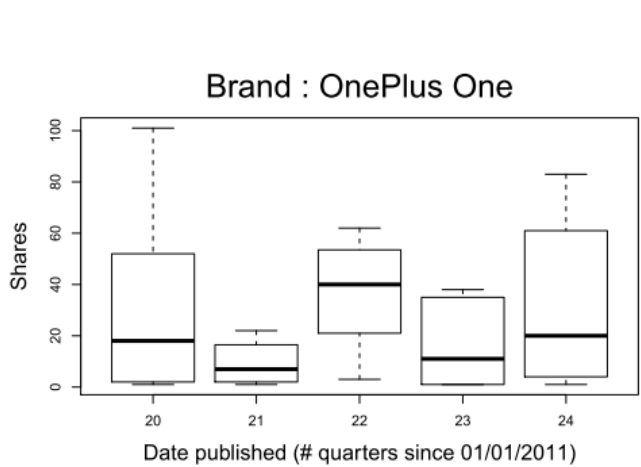

*FIGURE 21.* SHARES BY DATE PUBLISHED

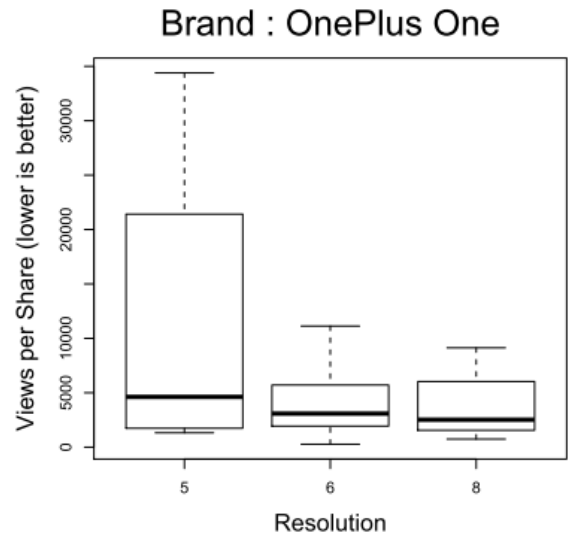

*FIGURE 22.* VIEWS-PER-SHARE BY RESOLUTION

The impact of *OnePlus One* (the first OnePlus phone) videos kept improving over time. These findings are crucial to our research to understand the movement of trend between the old and the new video attributes as the video production techniques improve. Apart from the purpose of balancing the results, we decided to include *OnePlus* for the following three reasons: (a) *OnePlus* did not use traditional media for advertising and marketing (Rosenstein, 2015). Therefore, almost all of the promotion for *OnePlus One* was through online reviews, blog conversations and online campaigns (b) *OnePlus* is considered to be a *'flagship killer'* (Kumar) because it was designed to be technologically and architecturally competitive (in some aspects, superior) to the flagship products by other popular smartphone companies, yet sold at a comparatively lower price (c) *OnePlus* was available by invitation only. We found that, OnePlus One's performance in terms of views-per-like improved over time as reviewers adopted higher resolution (See Figure 19) technology. Figure 22 shows improved views-per-share of OnePlus One reviews as resolution of the videos improved. Therefore it was interesting to further study the video attributes, especially shares and comments, of the video reviews on *OnePlus* because the brand seemed to generate curiosity and brand popularity among the technologically sound audience. Figures 20 and 21 show that the likes and shares of OnePlus One videos increased significantly during the 22$^{nd}$ quarter and remained fairly high afterward. Figure 22 shows an almost consistent views-per-share by resolution. Additionally, Figure 23 shows consistent video duration across time. This indicates that the brand has a well-crafted branding/marketing strategy.

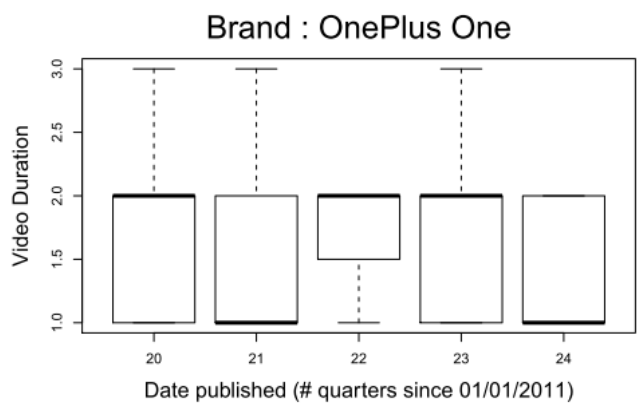

*FIGURE 23.* VIDEO DURATION BY DATE PUBLISHED

**Brand Performance Correlation Model**

The graphs explained above provide the intuitive understanding of dependent and the independent variables in the statistical model that is presented in this section. For example, let's consider the potential correlation between *views-per-like* (dependent variables) and three different independent variables: Resolution, Video duration, and Date published. In the following model, we statistically present the following findings using the technique of Linear Regression through the programming language R (Commonly used by Statisticians). R provides a built-in method that implements linear regression-based model. It accepts, as input, various dependent and independent data points and outputs the following:

1. The correlation and statistical significance between independent variables and dependent variables.

2. Coefficients and an error constant (e) to derive dependent variables using the independent variables.

The input that we provided to the linear regression model was all data points/brands except *Nokia* and *BlackBerry*. Based on the previous chapter, restating the output of the linear regression model in R is:

1. Formula: To establish a correlation between views-per-like (dependent variables) and the independent variables.

2. Residuals: Best-case error between the actual and the fitted values of views-per-like

3. Coefficients: These are the constants (value that do not change) that give you the best case estimation using the model.

**Formula.** Views/likes == *a* x tonality + *b* x resolution + *c* x video duration (seconds) + *d* x date published + *e*

a, b, c, d, and e (intercept) are constants generated by the linear regression model.

Residuals (Difference between estimated and actual views per like):

| Minimum | Median | Max |
|---|---|---|
| -157.38 | -16.71 | 810.49 |

**Meaning.** This is the range of difference between actual and modeled values for views-per-like. Minimum is the worst case underestimation of *views-per-like*. Max is the worst case overestimation of *views-per-like*. Median value is the common case estimation; in this case the estimation may be incorrect by -16.71 *views-per-like*. The minus is overestimation.

TABLE 14

*PREDICTION MODEL*

| Term | Coefficient (e, a, b, c, d) | Legend |
|---|---|---|
| Intercept | -6.325e+01 | ** = Views per like moderately correlated to Video Duration |
| Tonality | -1.051e+00 | *** = Views per like highly correlated to Date Published |
| Resolution | -4.190e+00 | **Meaning of 'e'** |
| Video Duration | 1.565e+01** | 6.326e+01 (value) = 6.32 * $10^1$ |
| Date Published | 5.45e-03*** | 6.326e+02 (value) = 6.32 * $10^2$ |
|  |  | 6.326e+03 (value) = 6.32 * $10^3$ |

*Note:* If the value is negative to the left, the right will also be negative.

So based on the formula, the model becomes,

Estimated *views-per-like* = (-1.051e+00) x tonality + (-4.190e+00) x resolution + 1.565e+01 x video duration + 5.456e-03 x date published + (-6.326e+01)

**Prediction.** We use the above model to predict *views-per-like* for videos that include reviews on *BlackBerry and Nokia*. We generated two plots (for BlackBerry and Nokia) each of which shows the predicted (blue) and actual (red) *views-per-like* for all videos that were studied.

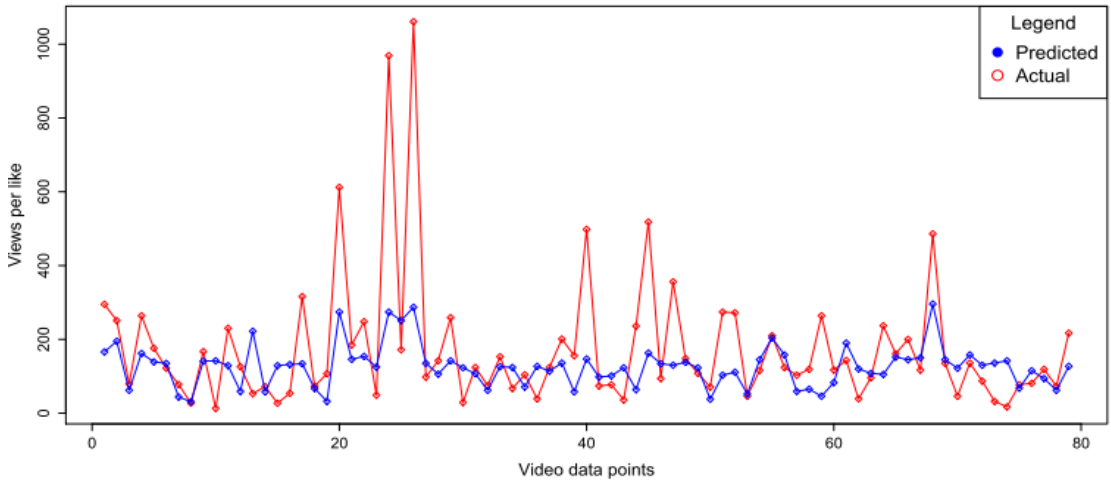

FIGURE 24. VIEWS-PER-LIKE BY VIDEO DATA POINTS FOR BLACKBERRY

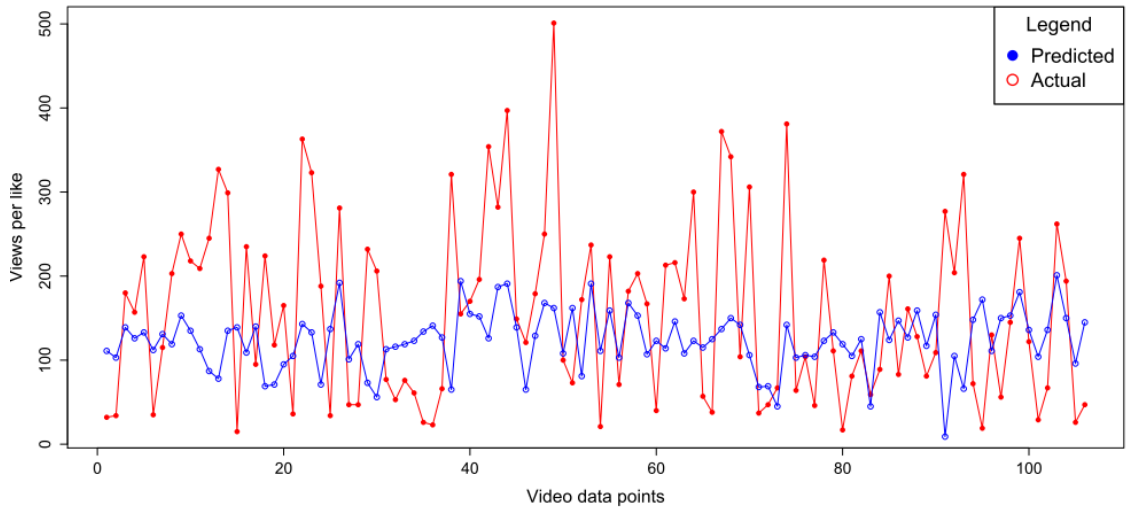

FIGURE 25. VIEWS-PER-LIKE BY VIDEO DATA POINTS FOR NOKIA

As seen in Figure 24 and Figure 25 actual vs predicted, the model is able to capture the trend in *views-per-like*, and the predicted values follow the actual values in majority of the cases. In fact, in case of some videos the model is accurately able to predict the actual *views-per-like* (See Figure 24). Recall that the model was constructed based on all videos except *Nokia* and *BlackBerry*. This shows two interesting findings: First, as seen in Figure 24, the graphs shows that the model is useful in predicting *views-per-like* of a given video review. Second, the model is a useful tool to improve future performance of video reviews.

However, as seen in Figure 25, the model is not able to predict the trend as effectively as for BlackBerry. Therefore, the graph for Nokia suggests that a model that works for a particular brand may not work for another brand even if the reviewer and attributes of the video remain the same. Hence, this indicates that a model could be constructed for individual brands instead of constructing one model for all brands combined.

With our method, brand managers can leverage the services of the reviewers to generate high audience attention. To prove this, we validated the method on another platform for a different industry. As discussed earlier, we used the Linear Regression method on a data set collected through Yelp.com for restaurant industry reviews.

While we considered views-per-like as our parameter to analyze the impact of YouTube reviewers on the reaction of their audience, we looked at words used in the Yelp reviews as a reaction to performance by restaurant staff as they are the carriers of the brand's message and hence contribute in brand perception. We considered the correlation between numbers of words i.e. wordcount and the number of occurrences of positive (or negative to support the analysis, similar to considering dislikes for YouTube videos in some cases) words similar to views-per-like. We based our assumption by drawing similarities between a restaurant visit and viewing a video on YouTube.com. We observed that, writing positive about the outlet is similar to like, comments, or shares for the outlet. In order to cover a very specific data, we based our investigation on areas highly dominated mostly by the millennial. We studied Tex-Mex outlets in a 10-mile radius around selective densely populated college campuses in the United States. For this research, we considered the following parameters:

1. Wordcount-per-positive-word (also covers negative words): The correlation between the number of words used and the occurrences of positive or negative words. Less wordcount and more positive words is an indication of a positive reviews towards the restaurant. Likewise, less negative words in more wordcount is a good indication of less complaints about the restaurant. We consider these metrics to measure performance of the restaurant also in terms of food and service.

2. Days/Quarter: This represents a moment in the plot i.e. it calculates the number of days required for an addition positive or negative word generated through the review. We study the gap between two data points for the days to calculate average number of days required to receive one extra positive word or negative word. This show a direct reaction to the performance of the outlet within specific states. Lesser the gap, better the response of the audience.

# Results

**Wordcount-per-Positive-Word by Quarter**

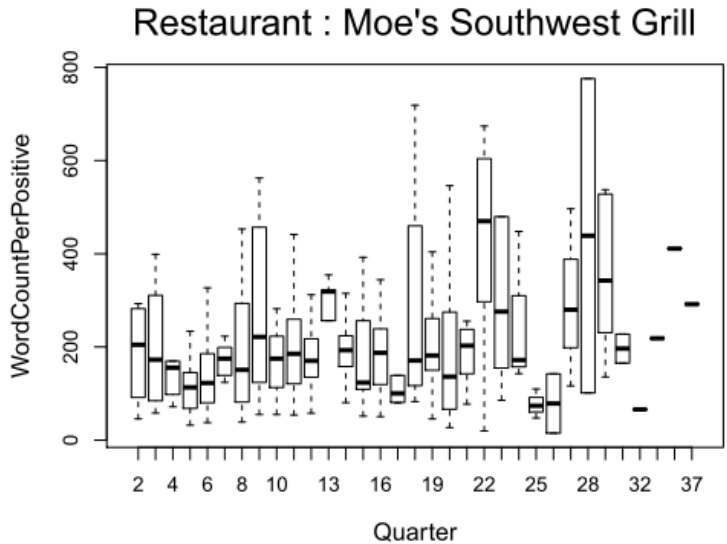

*FIGURE 26.* WORDCOUNT-PER-POSITIVE BY DATE
PUBLISHED IN QUARTER – PER RESTAURANT

*Wordcount-per-positive* is the number of positive words generated compared to the total number of words in the review. We study the correlation between wordcount-per-positive and days measured in set of Quarters. High number of positive words in a low wordcount represents a positive response towards the brand. We collected the data from 2006 to January 2016 (similar to date published in the smartphone analysis). As we observe in Figure 26, Moe's had a good review rate in the beginning until the 22$^{nd}$ quarter when wordcount-per-positive-words increased and remained higher compared to the set before the 22$^{nd}$ quarter. This shows that either Moe's performance in terms of service and food diminished over time or they did not get the required feedback from their audience in some states. To verify, we examined Wordcount-per-positive by. Quarter for each state (consider HI and CT) for Moe's Southwest Grill. We observed that in the 22$^{nd}$ quarter backwards from January 2016 i.e. in the year 2010 and 2011, Moe's Southwest Grill generated 27 reviews in Connecticut (Figure 27) and 22 reviews in Hawaii (Figure 28) which were more than other states. Observe that the wordcount-per-positive-words is high, which means that the spread of positive words were more throughout the quarter. To further investigate, we looked at wordcount-per-negative by Quarter for outlets in Connecticut and Hawaii for Moe's Southwest Grill. Observe in the Figures 29 and 30, the wordcount-per-negative is low, which means that the reviewers talk negatively about a brand in few words. This is extremely hurtful for any company since the words used in the review are crisp and can be remembered by the reader of the review for a long time. This will hurt any company most specifically a company in an industry like restaurant where most of the word spreads through people. To support these claims, we manually read the number of wordcount-per-positive-words and number of wordcount-per-negative-words for reviews in 2010 and 2011 for Moe's Southwest Grill, Hawaii.

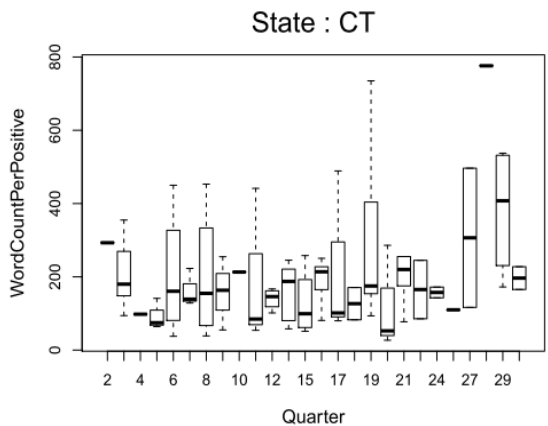

*FIGURE 27.* WORDCOUNT-PER-POSITIVE BY DATE PUBLISHED IN QUARTER FOR CT

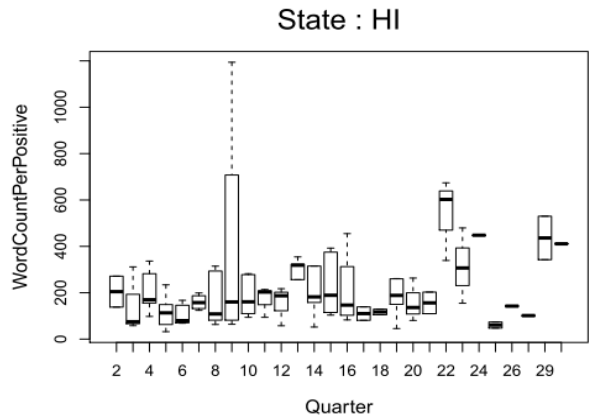

*FIGURE 28.* WORDCOUNT-PER-POSITIVE BY DATE PUBLISHED IN QUARTER FOR HI

We found that while the positive comments covered reviews about food and value for money, negatives comments were focused on customer service. Here is an example of a Yelp.com review by Jojo T:

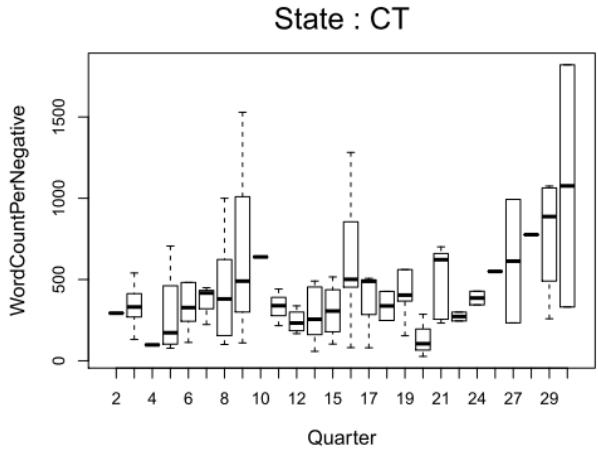

*FIGURE 29.* WORDCOUNT-PER-NEGATIVE BY DATE PUBLISHED IN QUARTER FOR CT

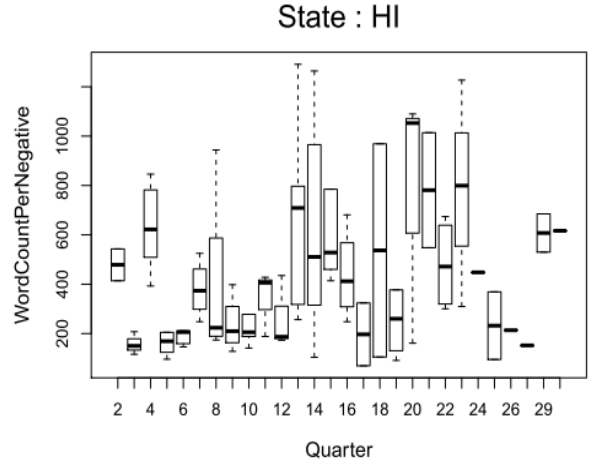

*FIGURE 30.* WORDCOUNT-PER-NEGATIVE BY DATE PUBLISHED IN QUARTER FOR HI

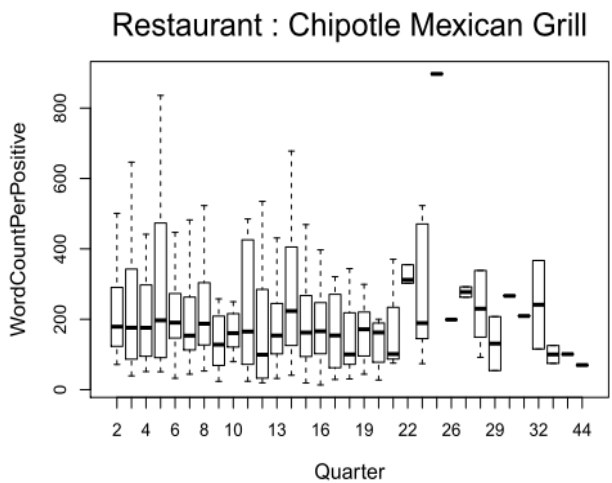

*FIGURE 31.* WORDCOUNT-PER-POSITIVE BY DATE PUBLISHED IN QUARTER – PER RESTAURANT

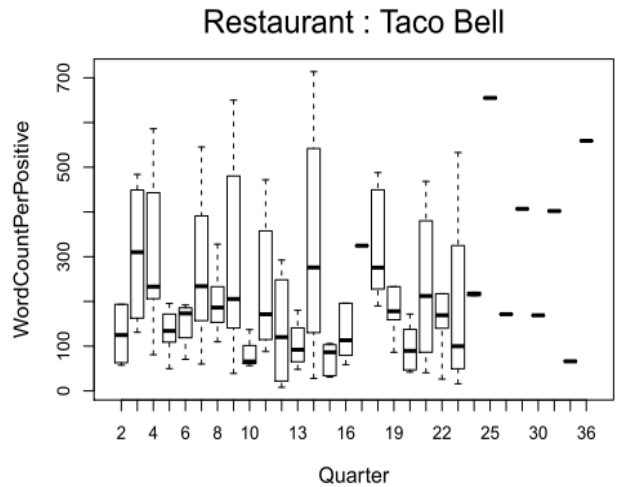

*FIGURE 32.* WORDCOUNT-PER-POSITIVE BY DATE PUBLISHED IN QUARTER – PER RESTAURANT

"Both times I went there: customer service was horrible!!!!! Seriously they need to retrain their staff. They always seem so grumpy and I felt like I was a bother to them. Food wasn't bad: better than Taco Bell. I ordered a beef and chicken burrito to go. The grouchy lady named TABITHAH helped me. Since it was a take out order: she dumped chips in a paper brown bag on top of my burritos. That's so ghetto but oh well. Main thing is that I'm never going back unless they drop their BAD ATTITUDES!!!!!!!!!!!!!!!!!!!!!!!!!!"

Observe that Jojo T was clearly disappointed by the customer service while he was okay with the food as he compared it with Taco Bell. It is clear from this example that the number of words used in the criticism were not only high, but also clearly written, which certainly is harmful to the reputation of the company especially because Jojo's 130 Yelp friends were notified about his review.

Given this situation, a brand manager's next step need to be towards (a) avoiding further harm to the brand perception by replying directly to Jojo (on Yelp) and (b) following-up with the performance of the outlet staff since they are the carrier of the brand message. Our method will enable brand managers to keep a track of the most important reviews and respond to sustain a good brand experience.

A successful brand can be identified by high number of positive words in lesser number of word count. We present two graphs to show the best and worst case examples. A good example is Chipotle Mexican Grill. Observe, in Figure 31, Chipotle Mexican Grill has a consistent performance. In contrast, as seen in Figure 32, Taco Bell shows inconsistency in the correlation between the spread of wordcount-per-positive-words through all quarters from 2007.

The next section discusses future work in this area and offers a conclusion.

# CHAPTER 5

# FUTURE WORK AND CONCLUSION

## Future Work

This research can be expanded locally as well as globally. For this thesis, we considered YouTube content creators based in United States and Canada. Prospective researchers can apply our method to find the impact of content creators on brand perceptions located in countries other than United States and Canada. Our research does not categorize the response by audience based on specific locations within United States and Canada for smartphone reviews, it will be interesting to identify (and probably categorize) specific locations of commenters (audience) to measure the reach of the content creators especially in the technology industry. Our research also does not cover views-per-share since we restricted our study only on data generated through YouTube.com and Yelp.com and not the one carried forward to other social media platforms. Sharing a YouTube video or Yelp review is an advanced analysis since it gives an opportunity for researchers to measure the impact of YouTube and Yelp sharing on the audience engagement through other platforms. For example, further research can be conducted on measuring audience engagement on a YouTube video through measure the impact of sharing it on Facebook.com. We also believe that as YouTube grows in popularity, the advancement of product comparison videos will be more significant. These videos are already a part of the YouTube data set but they are not yet popular among YouTube users. Researchers/brand managers can use our method to find the impact of content creators covering a comparison between two products/brands on the response of the audience.

On Yelp.com, our method will also benefit researchers and brand managers to understand the correlation between reviews and star ratings of those reviews.

## Conclusion

Our research analyze the impact of content creators on the audience engagement through YouTube.com and Yelp.com. In this thesis, we quantify specific attributes or YouTube video production techniques used by content creators like resolution, tonality, date published, and duration of the video to understand their impact on views-per-like for those videos. Our study is based on smartphone reviews by content creators in the United States & Canada. We prove that views-per-like is an important metric to understand the audience engagement through YouTube videos to its full potential. Views-per-like is a metric that quantifies the number of views required to receive one like for a specific video. Our major contribution is a prediction model based on these attributes and metrics for six smartphone companies.

We validate our method by applying it to the restaurant industry specifically four Tex-Mex outlets located in a 10-miles radius of densely populated (randomly selected) university campuses in the United States. Just like the content creators are carriers of the brand message in the smartphone industry, the staff working at these restaurant outlets are carriers of the brand message for popular Tex-Mex restaurants. Our research quantifies the impact of this brand message delivered through food and service on the response of the audience through Yelp.com reviews. While views-per-like is an important parameter for YouTube.com study, we found that wordcount-per-positive or wordcount-per-negative is an important measurement metric for Yelp.com reviews. This metric analyze a positive or negative word in the total number of words of the review. Through Yelp.com, we presented our next contribution which was the validation of our method used on YouTube.com reviews. We found that with minor changes in the attributes and metrics used to quantify Yelp.com data, we can help brand managers get real-time feedback from the audience on their products.

Our research is useful for brand managers and marketers to most effectively use social media content creators and impact the brand perception in the minds of their current and prospective customers/target audience. This will also enable companies to experience high level of interaction with their audience and an enhanced online promotion orchestrated by content creators and carried forward by their target audience.